\documentclass[mathleft,fleqn,%
 final,%
]{an}
\usepackage{graphicx}
\usepackage[varg]{txfonts}
\usepackage{pdflscape}
\usepackage{natbib}
\bibpunct{(}{)}{;}{a}{}{,}
\newcommand{\mygi}{MyGIsFOS}
\begin{document}

\Pagespan{1}{}
\Yearpublication{2017}%
\Yearsubmission{2017}%
\Month{0}%
\Volume{999}%
\Issue{0}%
\DOI{asna.201400000}%

\title{The {\it Pristine} survey II: -- \\
        a sample of bright stars observed with FEROS   \thanks{Data
from FEROS}}

\author{E. Caffau\inst{1}\fnmsep\thanks{Corresponding author:
        {Elisabetta.Caffau@obspm.r}}
\and  P.~Bonifacio\inst{1}
\and  E.~Starkenburg\inst{2}
\and  N.~Martin\inst{3,4}
\and  K.~Youakim\inst{2}
\and  A.~A.~Henden\inst{5}
\and  J.~I.~Gonz\'alez Hern\'andez\inst{6,7}
\and  D.~S.~Aguado\inst{6,7}
\and  C.~Allende~Prieto\inst{6,7}
\and  K.~Venn\inst{8}
\and  P.~Jablonka\inst{9,1}
}
\titlerunning{Pristine II}
\authorrunning{E. Caffau et al.}
\institute{
GEPI, Observatoire de Paris, PSL Research University, CNRS, Place Jules Janssen, 92190 Meudon, France
\and 
Leibniz-Institut f\"ur Astrophysik Potsdam (AIP), An der Sternwarte 16, 14482 Potsdam, Germany
\and 
Universit\'e de Strasbourg, CNRS, Observatoire astronomique de Strasbourg, UMR 7550, F-67000, France
\and
Max-Planck-Institut f\"ur Astronomie, K\"onigstuhl 17, D-69117 Heidelberg, Germany
\and
AAVSO, Cambridge, Massachusetts, USA
\and
Institudo de Astrof\'isica de Canarias, V\'ia L\'actea, 38205 La Laguna, Tenerife, Spain
\and
Universidad de La Laguna, Departamento de Astrofisica, 38206 La Laguna, Tenerife, Spain
\and
Dept. of Physics and Astronomy, University of Victoria, P.O. Box 3055, STN CSC, Victoria BC V8W 3P6, Canada
\and
Laboratoire d'Astrophysique, Ecole Polytechnique F\'ed\'erale de Lausanne (EPFL), Observatoire de Sauverny, CH-1290 Versoix, Switzerland
}

\received{03.2017}
\accepted{05.2017}
\publonline{05.2017}


\abstract{%
Extremely metal-poor (EMP) stars are old objects formed in the first Gyr of the Universe.
They are rare and, to select them, the most successful strategy
has been to build on large and low-resolution spectroscopic surveys.
The combination of narrow- and broad band photometry provides a powerful and cheaper
alternative to select metal-poor stars.
The on-going Pristine Survey is adopting this strategy, conducting photometry with 
the CFHT MegaCam wide field imager and a narrow-band filter centred at 395.2\,nm on the \ion{Ca}{ii}-H and -K lines. 
In this paper we present the results of the spectroscopic follow-up conducted on
a sample of 26 stars at the bright end of the magnitude range of the Survey ($g\la 15$), 
using FEROS at the MPG/ESO 2.2\,m telescope. 
From our chemical investigation on the sample, we conclude that this magnitude range is too bright to use the SDSS $gri$
bands, which are typically saturated. Instead the Pristine photometry can be usefully
combined with the APASS $gri$ photometry to provide reliable metallicity estimates.
}

\maketitle

\section{Introduction}

Extremely metal-poor (EMP, [Fe/H]$\le -3$) stars are old objects that formed from a gas cloud that did
not yet have the time to be enriched in metals by supernovae explosions,
the last stage of the evolution of the massive stars.
This low metal-content of the gas was the typical chemical composition of the pristine Universe,
at redshift $ z > 5$, i.e. more than 12.5\,Gyr ago \citep[see e.g.][Figure 2]{Pallottini}.
Among the stars formed from metal-poor clouds at this early stage of the Universe, 
only those with low mass (mass lower than the Solar mass) are still
observable today because their life-time is longer than the age of the Universe.
Stars more massive than the Sun had time to evolve and at present they are underluminous
compact objects:  neutron stars
or black holes, remnants of type II supernovae explosion for the most massive stars, 
or  white dwarfs for the less massive stars.
The EMP stars are very rare objects and, in order to find them, large amounts of data have to be gathered and analysed.
Several projects focused on the search and chemical investigations of EMP stars
are based on the analysis of low-resolution spectra obtained by large surveys 
in order to select the most promising
candidates to observe at medium- or high-resolution \citep[see e.g.][]{BPS85,BPS92,christlieb08,topos1}.

Observing stars  photometrically is much faster than to take spectra; 
one can also go deeper and observe fainter objects with the same telescope size and integration time.
A classical metallicity-sensitive colour is $U-B$, that however saturates
its sensitivity at metallicity around --2.0. Recently \citet{Monelli13} have introduced the
index $C\_ UBI = (U-B)-(B-I)$ that turned out to be very useful to identify
multiple populations in Globular Clusters and dwarf galaxies \citep{Monelli14}.
\citet{Fabrizio} have shown that this index is also metallicity sensitive and
one may expect similar performances also for analogous $C\_ ugr$ or $C\_ ugi$ colours.
Yet their sensitivity still has to be tested at metallicities below --3.0.

The spectra of EMP stars are characterised by a small number of weak metallic absorption lines.
Even the strongest metallic absorption lines in the Solar spectrum disappear or become extremely weak in
the spectrum of an EMP star. When a low-resolution spectrum (resolving power of: $R \approx$\,2000) of a metal-poor star is analysed,
the lines belonging to the \ion{Mg}{i}b and the infra-red \ion{Ca}{ii} triplets are usually detected, but in the
EMP regime these lines are usually not detectable. The only feature that is always present 
is the strong \ion{Ca}{ii}-K absorption line at 392\,nm.
This line is sometimes the only metallicity indicator to select EMP candidates from low-resolution spectra.
This feature is so strong that its presence is also detectable in photometric observations.
To make the signature of this line clear, narrow-band photometry centred at about 395.2\,nm is the best solution
and EMP candidates can be efficiently and reliably selected.

The use of a narrow band filter centred on the \ion{Ca}{ii} H and K lines, in conjunction with Str\"omgren
intermediate bands, as a means to obtain metallicity estimates down to very low metallicities is well
established \citep[][and references therein]{AT91,TAT95,AT98,T07}.
The {\it Pristine} Project \citep{pristine1} 
extends this technique by coupling a narrow band \ion{Ca}{ii} H and K filter
(CaHK) to the broadband survey $gri$ photometry, e.g. the SDSS bands \citep{sdss00}.
To do this it uses the wide-field imager MegaCam \citep{MegaCam} mounted
on the Canada France Hawaii Telescope (CFHT). 
We are performing follow-up spectroscopy with several telescope/spectrograph
combinations at both low ($R\sim 2000$) and high ($R>40000$) spectral resolution.
The spectra will allow us to derive the metallicities of the stars in order to better 
calibrate the photometric indices (see Youakim et al. in prep.).
We present here the chemical analysis of a sample 
of 26 among the brightest objects that have been observed with FEROS \citep{feros98,feros04}
at the MPG/ESO 2.2\,m telescope at La Silla.

\section{Observations and data reduction}

The MPG/ESO 2.2\,m FEROS data were taken during a visitor mode observing run from April 13 to 16, 2016. 
They were automatically reduced using the FEROS pipeline \citep{kaufer2000}. 
Typically, three sub-exposures were taken per object and these are combined using the IRAF \emph{combine} task 
using an average sigma clipping rejection. Any pixels which still lie more than 3$\sigma$ from the continuum 
in the combined spectrum are clipped to remove remaining cosmic rays. 
FEROS observes with a simultaneous sky fiber and we follow the same procedure for these spectra before 
subtracting the sky spectrum from the object. 

Finally, the object spectrum is shifted to rest wavelengths coss-correlating a synthetic spectrum for the \ion{Mg}{i}b region 
of a typical EMP giant, which has been created using an (OS)MARCS stellar atmosphere and the 
Turbospectrum code \citep{alvarez98,Gustafsson08,Plez08}.

\section{Calibration of the HK filter on the AB system}

The concept of AB magnitude was introduced by \citet{OkeGunn83}
who defined that magnitude zero \relax in any band corresponds to the magnitude of an ideal body
with constant flux density at any wavelength, such that 
$f^0_\nu =  \rm 3.631\times 10^{-23} W Hz^{-1} m^{-2}$.
The SDSS $ugriz$ system is such an AB system. 

If we denote as $S_\lambda$ the instrument response
function of our system (including filter transmission, 
transmission of the system optics and quantum efficiency
of the detector)  and $f_\lambda$ the flux of the object
to be measured we define 
\begin{equation}
m_{AB,S_\lambda} = -2.5\log\left(\int \lambda f_\lambda S_\lambda d\lambda\right)
+2.5\log\left(\int \lambda f^0_\nu S_\lambda d\lambda\right) 
\end{equation}

The second term of the right hand-side of the equation ensures that 
$m_{AB,S_\lambda}$ is zero for the ideal constant flux density
object and it can be easily computed taking into account that
$f^0_\lambda = f^0_\nu c /\lambda^2$, where $c$ is the speed
of light in vacuum.
We use as $S_\lambda$ the response function described in \citet{pristine1},
that is an average of the response function across the MegaCam field of view.
This takes into account the filter transmission, the quantum efficiency
of the CCD's and a model of the atmospheric transmission.
With this $S_\lambda$ the zero-magnitude constant is equal to $-9.9795$.

This definition is useful for synthetic photometry, however it does not
solve the practical problem of transforming instrumental magnitudes
to the standard system. This is achieved through observations of 
standard stars. The SDSS system defined 4 primary standard stars
\citep{Fukugita}, the magnitudes of these stars are derived by integrating
the flux given in that paper, multiplied by the SDSS instrument response function.
Table\,7 of \citet{Fukugita} provides the $ugriz$ magnitudes of the four
primary standards.

For the CFHT--MegaCam \ion{Ca}{ii} H and K filter, the magnitudes of the 4 primary 
SDSS standards are given in Table\,\ref{magstandard}.

\begin{table}
\caption{$CaHK$ magnitudes of the primary standard stars.\label{magstandard}}
\begin{tabular}{rrrr}
\hline
HD\,84937 & HD\,19445 & BD+26$^\circ$ 2606& BD +17$^\circ$ 4708\\
\hline
 8.7745 &   8.6615 &  10.2692 &  10.0545\\
\hline
\end{tabular}
\end{table}

In the SDSS photometry, the magnitudes of the primary standard stars
are used to transform the instrumental magnitudes observed with the
SDSS Monitor Telescope (60\,cm aperture) onto the standard system, these
are then tied to the secondary standards \citep{Smith}.
Efforts are under way to observe the SDSS primary standards with a
smaller telescope and a similar filter, that can be cross-calibrated with the
CFHT MegaCam CaHK filter. However, for the time being we do not
yet know precisely the absolute zero point of the CFHT $CaHK$ photometry.
We could not use the standard stars of the $hk$ system \citep{AT91,TAT95,AT98}
or the ones more recently established by \citet{Lee} and \citet{Calamida}
because they are all too bright to be observed directly with the CFHT 3.6\,m telescope.
In addition, since we are really interested in the colours obtained combining
the   $CaHK$ photometry with SDSS $gri$, the above-mentioned standard stars are 
too bright for SDSS as well. Finally, since the 
CaHK filters on the different telescopes are different, the use of the established
standards would require a careful cross-calibration, and likely also a dedicated
observational campaign.

We computed synthetic photometry from the fluxes of the grid of
ATLAS \citep{K05} models of \citet{CK03}, with the above-defined
zero point, we then applied a $-0.13$\,mag offset, as described in 
\citet{pristine1} to match the internal survey zero-point.

\section{Analysis}

\subsection{Selection from photometry}
\begin{figure}
\includegraphics*[width=77mm]{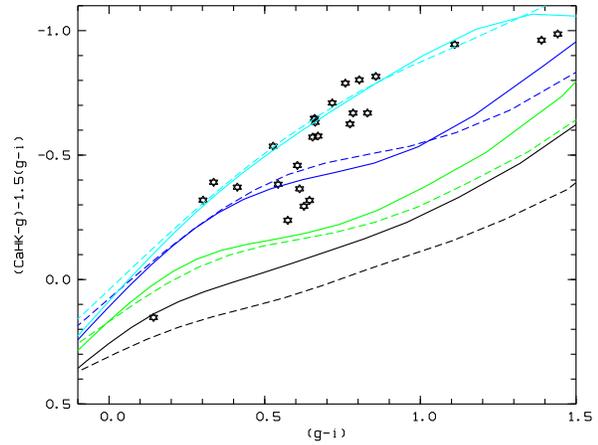}
\caption{The sample stars are represented as black stars.
The colours are derived from Pristine and SDSS photometry.
Solid and dashed lines represent the theoretical colours for $\log{\rm g}$ of 2.5 and 4.5, respectively.
Black lines are for solar metallicity, green lines for [Fe/H]=--1.0, blue lines for [Fe/H]=--2.0, light blue for [Fe/H]=--4.0.
}
\label{plotgisdss}
\end{figure}

\begin{figure}
\includegraphics*[width=77mm]{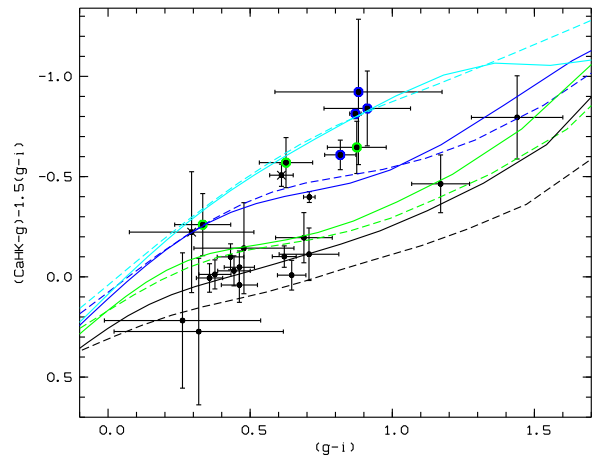}
\caption{The sample stars are  represented as black dots.
The colours are derived from Pristine and APASS photometry.
The stars with a larger blue dots are stars with $[{\rm Fe/H}]<-2.0$,
the green ones $-2.0<[{\rm Fe/H}]<-1.0$, [Fe/H] is from Table\,\ref{tlab}. 
Crosses represent spectra that have not been analysed.
Lines are as in Fig.\,\ref{plotgisdss}.}
\label{plotgi}
\end{figure}

The combination of the narrow-band filter CaHK and broad-band Sloan filters allows us to distinguish stars with
different metallicities, as shown in Figure\,3 of \citet{pristine1}.
In Fig.\,\ref{plotgisdss} a linear combination of broad-band Sloan filters and the CaHK as a function of the colour $(g-i)_0$ is 
compared to synthetic colours. 
From the plot, all stars in our sample, with the exception of one, are expected to be metal-poor, ${\rm [Fe/H]}<-1.0$.
However, all of these stars are at the bright-end for the SDSS Survey and they all have a flag of saturation for the individual filters.
We decided to observe them anyway because the global flag SDSS ``clean'' was set.
At the time of this first spectroscopic follow-up, SDSS photometry was the reference,
which led to the selection of stars. However, the first abundance investigation on this sample of stars 
revealed metallicities higher than expected. In order to understand this result, the first step was to check the photometry.
We then decided to use the $gri$ photometry of the APASS survey
\citep[][https://www.aavso.org/apass]{apass15,apass14,apass09} instead.
The colour-colour plot in  Fig.\,\ref{plotgi} is thus obtained:
the majority of the stars, which are metal-rich ([Fe/H]$>-1.0$) according to the chemical analysis described below (see Sect.\ref{chemi}), 
move down to a more appropriate place in the colour-colour space in the figure, 
close to the theoretical values corresponding to solar metallicity and
metal-poor stars are in the range of metal-poor theoretical colours.
The large uncertainties in the colours are dominated by the uncertainties in the APASS bands;
the CaHK band has a small uncertainty, negligible in the vertical error bars in Fig.\,\ref{plotgi}.

At the time we selected and observed the stars, the Pan-STARRS photometry \citep{Panstarrs} was not 
available. It is our plan to use this survey for the future star selection.

\begin{figure}
\includegraphics*[width=77mm]{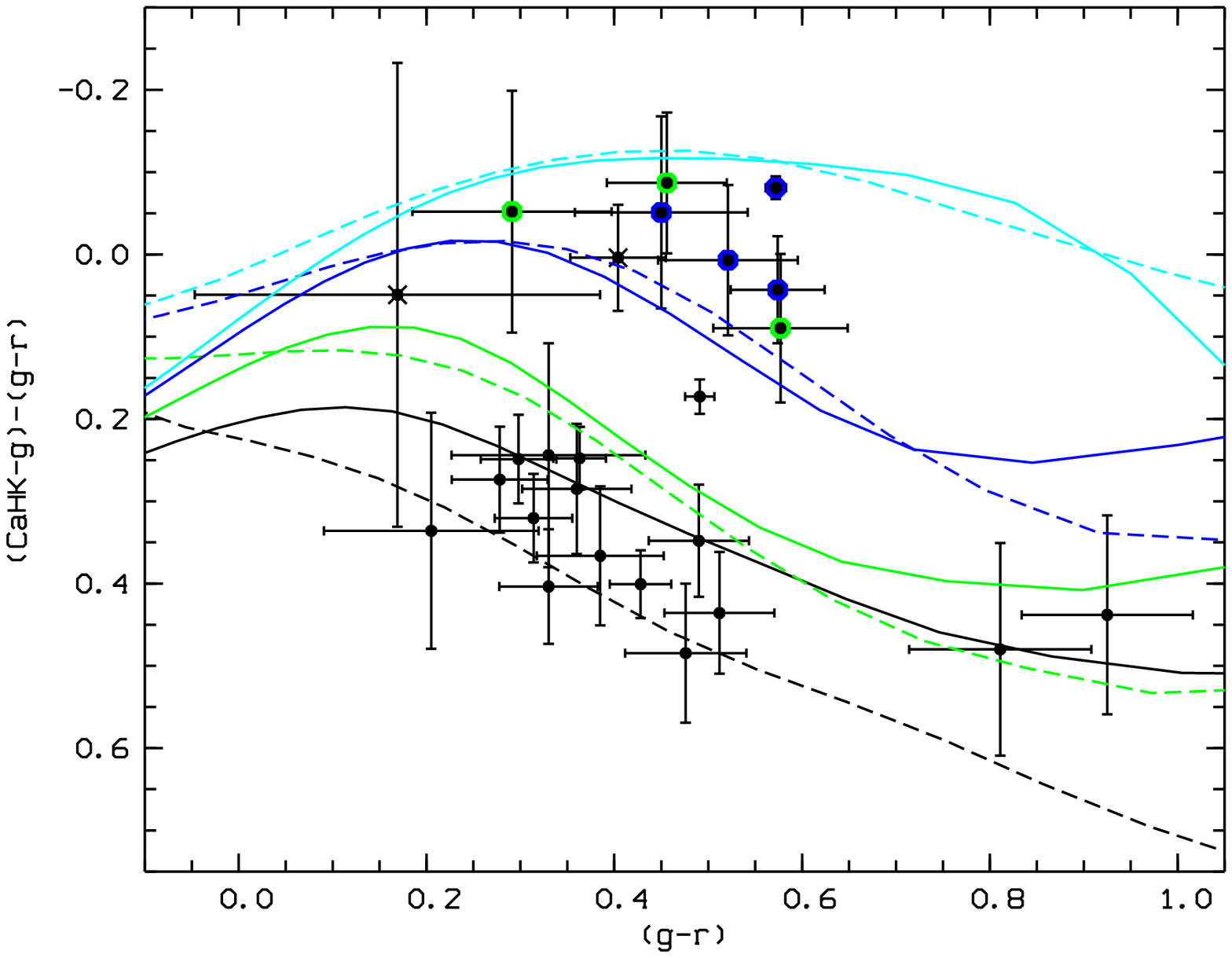}
\caption{The sample stars are represented as black dots.
The colours are derived from APASS photometry.
The stars with a larger blue dots are stars with $[{\rm Fe/H}]<-2.0$,
the green ones $-2.0<[{\rm Fe/H}]<-1.0$, [Fe/H] is from Table\,\ref{tlab}.
Crosses represent spectra that nave not been analysed.
Theoretical colour are as in Fig.\,\ref{plotgi}.
}
\label{plotgr}
\end{figure}

In Fig.\,\ref{plotgr}, we used the broad-band $r$ filter instead of $i$ and we can see a clearer separation
of metal-poor and solar-metallicity stars.
In addition in this case the metallicity derived from the spectra puts the stars in the plot closer to the corresponding
locus defined by theoretical colours.

\subsection{Chemical analysis\label{chemi}}
To derive the effective temperatures of the stars, taking advantage of the APASS filters, 
we used the calibrations from \citet{ivezic08}.
These authors provide the following expressions to derive ${\rm T}_{\rm eff}$:
\begin{enumerate}
\item $\log{\rm T_{\rm eff}}=3.882-0.316\left(g-r\right)_0+0.0488\left(g-r\right)_0^2+0.0283\left(g-r\right)_0^3$
applicable for $-0.3<\left(g-r\right)_0<1.3$;
\item ${5040\,{\rm K}\over {\rm T_{\rm eff}}}=0.532\left(g-r\right)_0+0.654$ in good agreement with expression (1).
\end{enumerate}

We derive the ${\rm T}_{\rm eff}$ from these two expressions. After iterations of the spectrum analysis, 
we used expression\,1 for unevolved stars and
expression\,2 for evolved stars. The absolute average difference among the two ${\rm T}_{\rm eff}$ scales 30\,K,
and 21 stars have a difference in the two ${\rm T}_{\rm eff}$ estimates smaller than 30\,K.
For the RR\,Lyr Pristine\_225.8227+14.2933 and for the cool dwarf Pristine\_255.0531+10.7488 the differences are slightly
larger than 100\,K.

With a fixed ${\rm T}_{\rm eff}$ we ran the code \mygi\ using a grid
of synthetic spectra computed with {\sf turbospectrum} \citep{alvarez98,plez12},
based on the grid of OSMARCS models and line-lists provided by the GES collaboration \citep{smiljanic14,heiter15}
and already used by \citet{duffau}.
\mygi\ is a code that analyses stellar spectra in order to derive the stellar parameters and the chemical composition
simulating a traditional abundance analysis; details can be found in \citet{sbordone14}.
HE\,1207-3108 has been observed in the same observation programme, in order to check the accuracy of our analysis.
The spectrum has a low signal-to-noise ratio (SNR of 25 per pixel at 500\,nm) so that
only few lines are detected in the spectrum of this metal-poor star.
We adopted the stellar parameters from \citet{yong13} (${\rm T}_{\rm eff}$/$\log{\rm g}$ of 5294\,K/2.85)
and a micro-turbulence of 1.0\,km/s, close to the 0.9\,km/s adopted by \citet{yong13}.
Analysing this spectrum as the others of the sample, we derive ${\rm [Fe/H]}=-2.78\pm 0.20$ to be compared to
${\rm [Fe/H]}=-2.70$ from \citet{yong13}.
We also agree with the result of \citet{yong13} on a low Ca abundance, but we can only detect two lines.
We also analysed the spectrum of HE\,1005-1439 that unfortunately has a low SNR of 10 at 500\,nm.
By adopting the stellar parameters from \citet{aoki07} 
(${\rm T}_{\rm eff}=5000$\,K, $\log{\rm g}=1.9$ and a micro-turbulence of 2.0\,km/s)
we obtain from two lines of neutral iron [Fe/H]$=-3.08\pm 0.12$, in good agreement with [Fe/H]=--3.17 of \citet{aoki07}.

\begin{figure}
\resizebox{77mm}{!}{\includegraphics[clip=true]{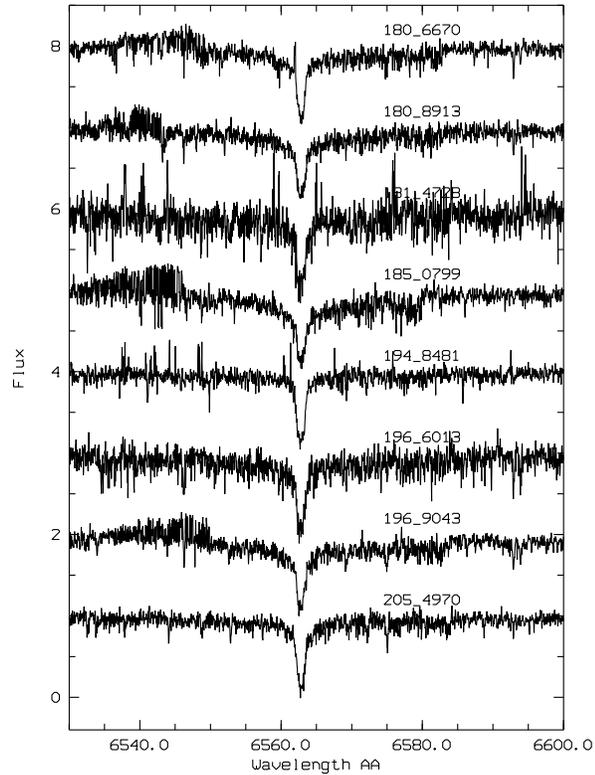}}
\caption{H$\alpha$ for eight of the analysed stars.}
\label{plotha1}
\end{figure}

\begin{figure}
\resizebox{77mm}{!}{\includegraphics[clip=true]{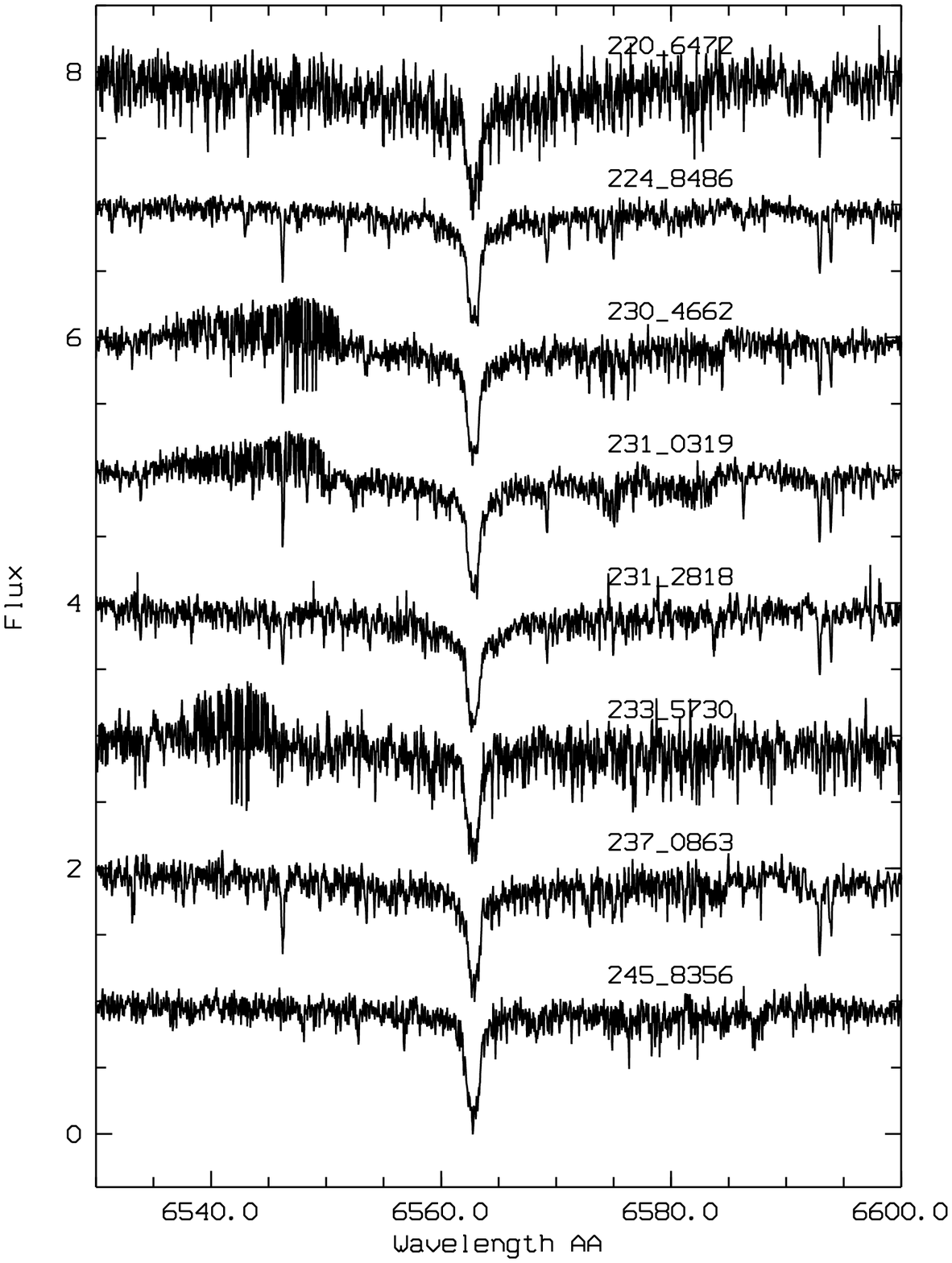}}
\caption{H$\alpha$ for the other eight of the analysed stars.}
\label{plotha2}
\end{figure}

\begin{figure}
\resizebox{77mm}{!}{\includegraphics[clip=true]{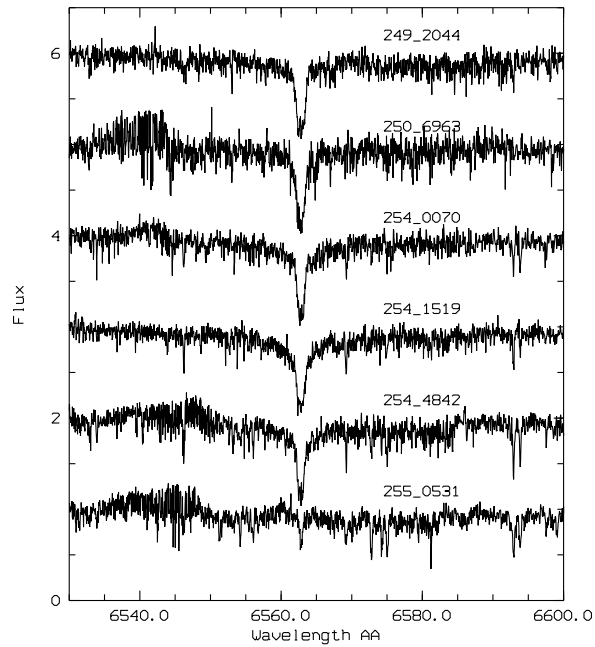}}
\caption{H$\alpha$ for the last six analysed stars.}
\label{plotha3}
\end{figure}

\begin{figure}
\resizebox{77mm}{!}{\includegraphics[clip=true]{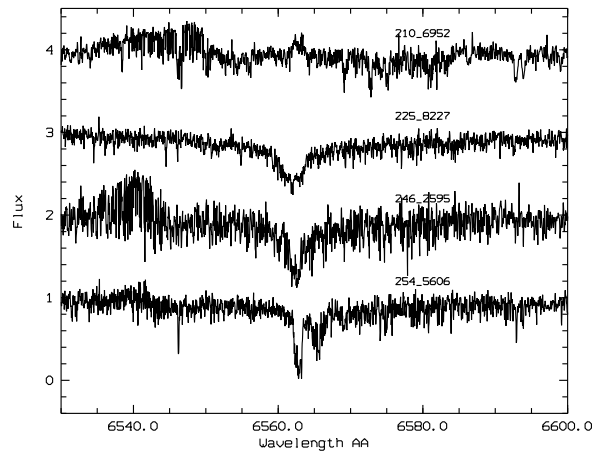}}
\caption{Four peculiar H$\alpha$ profiles.}
\label{plotha5}
\end{figure}

\begin{figure}
\resizebox{77mm}{!}{\includegraphics[clip=true]{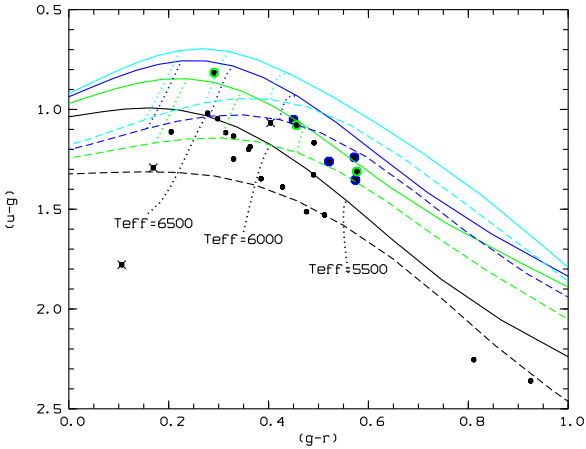}}
\caption{The $(u-g)$ vs. $(g-r)$ diagram for our program
stars. Overlayed synthetic colours for [M/H]=0.0 (black),
[M/H]=--1.0 (green), [M/H]=--2.0 (blue) and
[M/H]=--4.0 (cyan), for  log\,g\,=2.5 (solid lines) and
log\,g\,=4.5 (dashed lines).
[M/H] is the metallicity of the model, note that all the metal-poor models are
enhanced by +0.4\,dex in $\alpha$-elements.
}
\label{ug_gr}
\end{figure}

\begin{figure}
\resizebox{77mm}{!}{\includegraphics[clip=true]{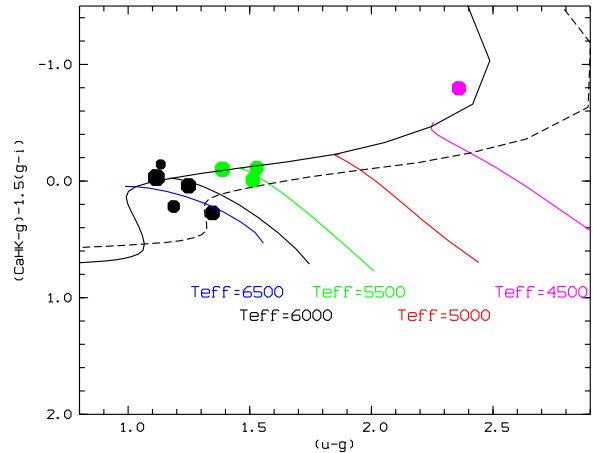}}
\caption{The $(u-g)$ vs.  $(CaHK - g) - 1.5(g-r)$   diagram for our program
stars with metallicity [M/H]$\approx 0.0$. 
The solid line corresponds to synthetic colours for log g = 2.5 
and the dashed lines to 
log g = 4.5. Each star is colour coded with the closest T$_{\rm eff}$
and the symbol size is proportional to  log g.
}
\label{ug_cak_m00}
\end{figure}

\begin{figure}
\resizebox{77mm}{!}{\includegraphics[clip=true]{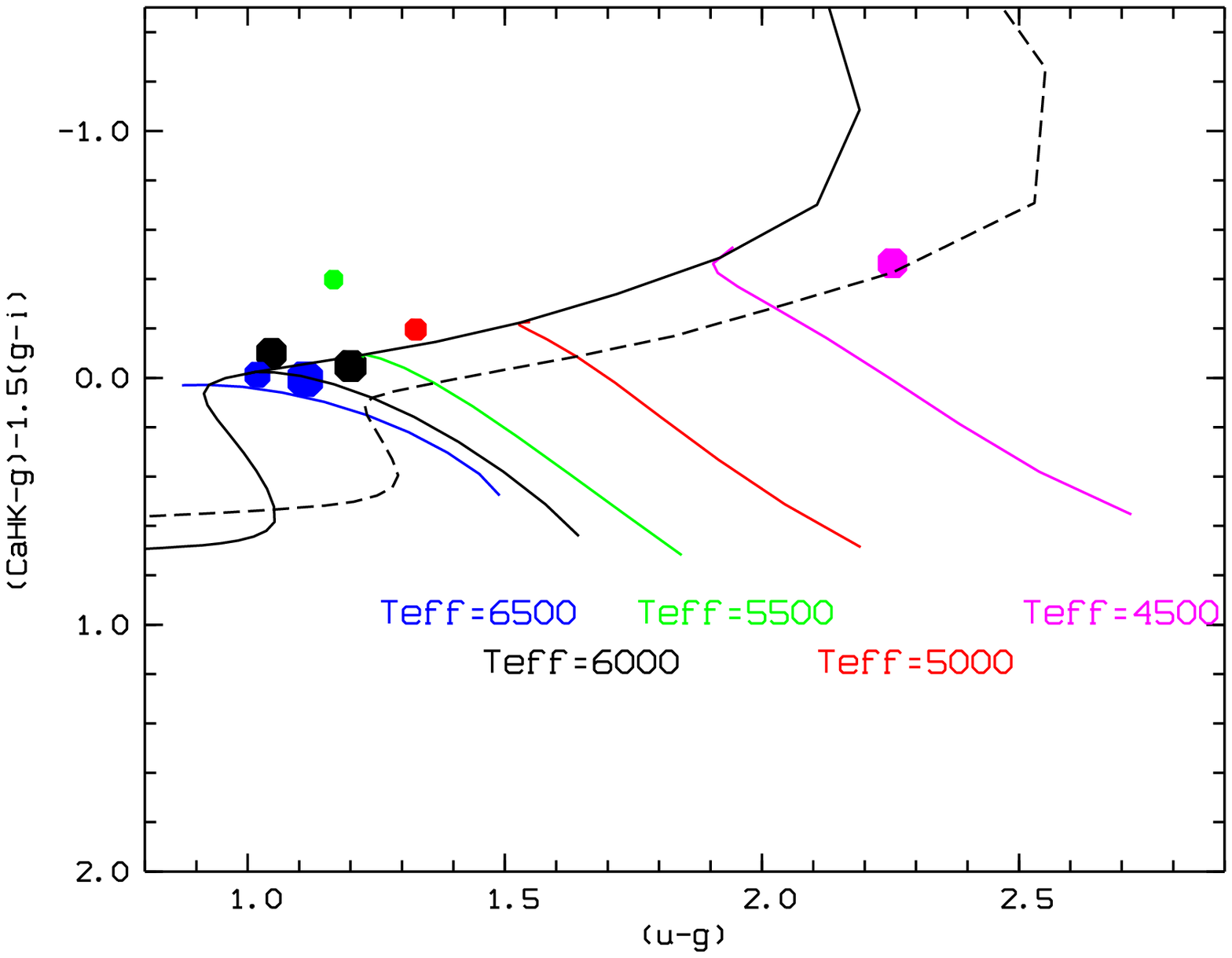}}
\caption{Like Fig.\,\ref{ug_cak_m00}, but for [M/H]=-0.5.}
\label{ug_cak_m05}
\end{figure}

For the stars in the sample, the surface gravity is based on the iron ionisation balance;
we also systematically compared the wings of the Mg\,I\,b triplet of the observed spectra 
to synthetic profile computed with the derived surface gravity as an extra-check on the surface gravity.
The micro-turbulence is derived from the agreement between the Fe abundance derived from weak and strong \ion{Fe}{i} lines.
For some spectra the signal-to-noise ratio is too poor to derive surface gravity
and micro-turbulence and we fix the stellar parameters as follows:
surface gravity looking only at the wings of the Mg\,I\,b triplet and the micro-turbulence using values of stars already known
and analysed in the literature with similar atmospheric parameters \citep[e.g.][]{santos05,ecuvillon04,pfr07}.

In Table\,\ref{tlab}, we provide the stellar parameters for the programme stars
included the barycentric radial velocities derived from our spectra.
Pristine\,225.8227+14.2933 is already known in the literature (SDSS\,J150317.44+141735.7) 
and is classified as an RR\,Lyr type \citep[see][]{drake13}.
We do not know at which phase this star has been observed and we cannot derive the effective
temperature from the photometry or the wings of H$\alpha$.
The quality of the spectrum is mediocre, with a signal-to-noise ratio (SNR) of 24 at 540\,nm.
We decided to look at the very few lines detected and we chose a particular couple of \ion{Fe}{i} lines
(at 540 and 543\,nm) whose depth ratio depends on the stellar temperature.
From this diagnostic we would derive a ${\rm T}_{\rm eff}$ of about 5400\,K.
But when we examined at the line-to-line scatter of the four \ion{Fe}{i} lines, the lowest
value is found for a ${\rm T}_{\rm eff}$ of about 6600\,K.
In Table\,\ref{tlab} we provide values for both cases.
We are confident that this star is metal-poor, with an iron abundance of the order of or below 1/100 the Solar value,
but we are not able to give a better determination.

Pristine\,246.2595+11.8378 is within $3''$ from CSS\,J162502.2+115017, also an RR\,Lyr type star \citep{drake13,Abbas14}.
With a poor spectrum quality, SNR of 10 at 540\,nm, we are not able to derive any 
information on the chemical composition.
Also the spectrum of Pristine\,181.4728+15.5306 is of poor quality and did not allow us to derive any chemical
information of the star.

The use of $hk$ index to derive metallicities
of RR\,Lyr stars has already been  demonstrated
both for the field \citep{Baird} and cluster  \citep{Rey} variables.
Our $CaHK$ photometry can provide similar performances. In our survey 
we expect to find EMP RR\,Lyr stars, such as are already known
to exist \citep{Wallerstein,Camilla} along with 
non-pulsating HB-stars \citep{Preston06,Seyma}.

\begin{table*}
\caption{Programme stars coordinates and atmospheric parameters. To derive [Fe/H] the iron Solar abundance is from \citet{abbosun}.}
\renewcommand{\tabcolsep}{3pt}
\tabskip=0pt
\footnotesize
\label{tlab}
\begin{tabular}{ccrcrccccl}\hline
Star                      & R.A.         & Dec.         & g$_0$ & V$_{\rm rad}$& Teff         & Logg & $\xi$ & [Fe/H] & Comment\\ 
                          & deg          & deg          & mag   & [km/s] & K            & cgs  & km/s  &  dex   & \\ 
\hline
Pristine\,180.6670+13.3324 &  180.6669617 & +13.33239746  & 13.9 & +30.0   & 6210  & 4.54 & 0.94 & $-0.60$ & A(Li)$_{\rm 1DLTE}=2.58$ \\ 
Pristine\,180.8913+11.3199 &  180.8912811 & +11.3199358   & 13.9 & +15.3   & 6290  & 4.28 & 1.00 & $-0.51$ & (Y), $\xi$ fixed \\ 
Pristine\,181.4728+15.5306 &  181.4727936 & +15.53064156  & 15.2 & --18.5  &       &      &      & $     $ & (Y), Only noise \\   
Pristine\,185.0799+14.6464 &  185.0799103 & +14.64640236  & 14.1 & +162.5  & 6240  & 3.24 & 1.50 & $-1.48$ & (Y), $\xi$ fixed \\  
Pristine\,194.8481+11.5875 &  194.8481293 & +11.58754444  & 14.7 & +119.7  & 5260  & 3.00 & 1.50 & $-2.64$ & (Y), log\,g and $\xi$ fixed \\ 
Pristine\,196.6013+15.6768 &  196.6013336 & +15.67680836  & 14.6 & --102.8 & 5970  & 4.63 & 1.00 & $-0.32$ & $\xi$ fixed \\    
Pristine\,196.9043+06.8973 &  196.9043121 &  +6.89729452  & 13.8 & --4.9   & 6080  & 4.01 & 0.86 & $-0.07$ & (Y), A(Li)$_{\rm 1DLTE}=2.60$ \\
Pristine\,205.4970+15.0564 &  205.4970245 & +15.05644989  & 14.4 & --13.5  & 5630  & 3.28 & 1.00 & $-1.59$ &  \\  
Pristine\,210.6952+12.8768 &  210.6952057 & +12.87676144  & 14.9 & --4.2   & 4640  & 4.50 & 1.00 & $-0.42$ & (Y), H$\alpha$ in emission \\ 
Pristine\,220.6472+15.7418 &  220.6472321 & +15.74179745  & 14.7 & +34.4   & 6600  & 4.82 & 1.59 & $-0.37$ &  \\  
Pristine\,224.8486+07.0259 &  224.8485565 &  +7.02591133  & 14.1 & +50.4   & 5730  & 4.64 & 1.04 & $-0.14$ &  \\  
Pristine\,225.8227+14.2933 &  225.8226929 & +14.29327202  & 13.5 & --93.7  & 6600  & 3.50 & 1.50 & $-1.87$ & (Y), RR Lyr \\ 
                          &  225.8226929 & +14.29327202   & 13.5 & --93.7  & 5400  & 3.50 & 1.50 & $-3.07$ & (Y), RR Lyr \\ 
Pristine\,230.4662+06.5251 &  230.4662018 &  +6.52514982  & 14.2 & --26.5  & 5520  & 3.93 & 0.21 & $-0.78$ &  \\ 
Pristine\,231.0319+06.4867 &  231.0318909 &  +6.48666382  & 14.2 & +22.0   & 5880  & 4.57 & 0.36 & $+0.03$ &  \\ 
Pristine\,231.2818+06.4018 &  231.2817535 &  +6.40175009  & 14.3 & --44.0  & 5960  & 4.30 & 1.27 & $-0.14$ &  \\ 
Pristine\,233.5730+06.4702 &  233.5729523 &  +6.47022772  & 14.7 & --79.2  & 5250. & 2.50 & 1.50 & $-2.28$ &  (Y) \\
Pristine\,237.0863+10.5790 &  237.0862732 & +10.57896423  & 14.3 & --28.1  & 5570  & 4.54 & 0.98 & $-0.09$ &  \\ 
Pristine\,245.8356+13.8777 &  245.835556  & +13.87771988  & 14.0 & --176.0 & 5650  & 3.44 & 1.00 & $-2.12$ & (Y), $\xi$ fixed \\
Pristine\,246.2595+11.8378 &  246.2595367 & +11.83775043  & 14.4 & +28.1   &       &      &      &         & (Y), RR Lyr \\
Pristine\,249.2044+10.5327 &  249.2044373 & +10.53268337  & 14.2 & --402.0 & 5240. & 2.07 & 2.00 & $-1.86$ &  \\ 
Pristine\,250.6963+08.3743 &  250.6963348 &  +8.37430286  & 14.6 & --4.0   & 5410  & 2.50 & 2.00 & $-2.12$ & (Y), log\,g and $\xi$ fixed \\ 
Pristine\,254.0070+12.7611 &  254.0070496 & +12.76109982  & 13.4 & --16.8  & 6080  & 4.63 & 1.58 & $-0.17$ &  \\ 
Pristine\,254.1519+12.6741 &  254.1518707 & +12.67414761  & 13.5 & +16.9   & 6140  & 4.82 & 0.30 & $-0.21$ &  \\  
Pristine\,254.4842+15.4573 &  254.4842224 & +15.4572525   & 12.8 & --8.1   & 5450  & 4.45 & 1.05 & $-0.08$ &  \\ 
Pristine\,254.5606+15.4784 &  254.5606079 & +15.47841072  & 13.8 & +24.2   & 5220  & 4.08 & 1.48 & $-0.67$ & (Y), double H$\alpha$ \& H$\beta$ \\ 
Pristine\,255.0531+10.7488 &  255.0530548 & +10.748806    & 14.7 & +82.1   & 4500  & 4.50 & 1.00 & $-0.17$ &  (Y) \\ 
\hline
\end{tabular}
\\
\smallskip
(Y) indicates stars that passed the selection criterion of the Pristine calibration.
\end{table*}

In Tables\,\ref{tlab_abun1} to \ref{tlab_abun3} the detailed chemical abundances derived from the spectra are provided.
In Fig.\,\ref{plotha1}, \ref{plotha2}, and \ref{plotha3} we present the quality of the spectra in the range of H$\alpha$.
In Fig.\,\ref{plotha5} the four peculiar stars: Pristine\,210.6952+12.8768, a star with H$\alpha$ emission, probably an active star,
the two RR\,Lyr stars and Pristine\,254.5606+15.4784, a double lined system. 
For Pristine\,210.6952+12.8768 we do not detect emission in H$\beta$. 
For Pristine\,254.5606+15.4784, we do not detect the lines of the secondary star
other than H$\alpha$ and H$\beta$. Even cross-correlation functions with synthetic
spectra do not show a secondary peak. We therefore conclude that the metallic lines of the secondary star
are too weak to be detected with our SNR ratio and we treat the star as single, assuming a negligible veiling.

For the two stars Pristine\,180.6670+13.3324 and Pristine\,196.9043+06.8973, we have a Li detection and abundances of
${\rm A(Li)}=2.58\pm 0.20$ and $2.60\pm 0.20$, respectively. When we apply the NLTE correction provided by \citet{lind09},
we derive ${\rm A(Li)}=2.53\pm 0.20$ and $2.58\pm 0.20$, respectively, for the two stars.
The presence of Li in metal-rich stars is not unusual, these stars are probably young Galactic stars.

\begin{table*}
\caption{Stellar individual abundances from Na to Sc as: 
$A\left({\rm X}\right)=\log_{10}\left({N\left({\rm X}\right)\over N\left({\rm H}\right)}\right)+12$.}
\renewcommand{\tabcolsep}{3pt}
\tabskip=0pt
\footnotesize
\label{tlab_abun1}
\begin{tabular}{lcccccccc}\hline
Star & \ion{Na}{i} & \ion{Al}{i} & \ion{Mg}{i} & \ion{Si}{i} & \ion{Si}{ii} & \ion{S}{i} & \ion{Ca}{i} & \ion{Sc}{ii} \\
\hline 
Pristine\,180.6670+13.3324 & $5.70\pm 0.18$ &                &                & $7.15\pm 0.14$ & $7.14\pm 0.30$ &                & $5.94\pm 0.11$ & $2.85\pm 0.14$ \\ 
Pristine\,180.8913+11.3199 & $6.00\pm 0.09$ &                &                & $7.18\pm 0.18$ & $6.99\pm 0.30$ &                & $5.93\pm 0.11$ & $2.88\pm 0.04$ \\ 
Pristine\,185.0799+14.6464 &                &                &                &                &                &                & $5.29\pm 0.19$ &                \\ 
Pristine\,194.8481+11.5875 &                &                &                &                &                &                & $3.91\pm 0.12$ &                \\ 
Pristine\,196.6013+15.6768 & $5.83\pm 0.22$ &                &                & $7.25\pm 0.30$ & $7.37\pm 0.00$ &                & $6.19\pm 0.19$ & $2.99\pm 0.09$ \\ 
Pristine\,196.9043+06.8973 & $6.11\pm 0.08$ & $6.54\pm 0.20$ &                & $7.44\pm 0.07$ & $7.46\pm 0.08$ & $7.34\pm 0.30$ & $6.24\pm 0.13$ & $3.24\pm 0.24$ \\ 
Pristine\,205.4970+15.0564 &                &                & $6.39\pm 0.30$ & $6.33\pm 0.00$ & $6.92\pm 0.30$ &                & $4.86\pm 0.04$ &                 \\
Pristine\,210.6952+12.8768 & $6.24\pm 0.07$ &                &                &                &                &                & $6.20\pm 0.00$ & $2.73\pm 0.20$ \\ 
Pristine\,220.6472+15.7418 & $5.80\pm 0.03$ &                & $7.55\pm 0.30$ & $7.53\pm 0.12$ &                &                & $6.28\pm 0.12$ & $3.25\pm 0.18$ \\ 
Pristine\,224.8486+07.0259 & $6.28\pm 0.02$ & $6.64\pm 0.15$ &                & $7.59\pm 0.20$ & $7.68\pm 0.30$ &                & $6.26\pm 0.03$ & $3.52\pm 0.13$  \\
Pristine\,230.4662+06.5251 & $5.30\pm 0.16$ &                &                & $6.95\pm 0.11$ &                &                & $5.93\pm 0.23$ & $2.63\pm 0.25$  \\
Pristine\,231.0319+06.4867 & $6.23\pm 0.09$ & $6.59\pm 0.23$ &                & $7.55\pm 0.13$ &                &                & $6.44\pm 0.19$ & $3.48\pm 0.13$  \\ 
Pristine\,231.2818+06.4018 & $6.20\pm 0.00$ & $6.51\pm 0.20$ &                & $7.52\pm 0.15$ & $7.63\pm 0.02$ &                & $6.22\pm 0.10$ & $3.13\pm 0.13$  \\
Pristine\,233.5730+06.4702 &                &                &                &                &                &                & $4.33\pm 0.12$ &                 \\
Pristine\,237.0863+10.5790 & $6.20\pm 0.19$ & $6.57\pm 0.23$ &                & $7.19\pm 0.15$ & $7.70\pm 0.30$ & $7.68\pm 0.30$ & $6.52\pm 0.00$ & $3.03\pm 0.16$ \\ 
Pristine\,245.8356+13.8777 &                &                &                &                &                &                & $4.01\pm 0.11$ &                \\ 
Pristine\,249.2044+10.5327 &                &                & $6.09\pm 0.30$ &                &                &                & $4.74\pm 0.22$ & $1.28\pm 0.27$  \\
Pristine\,250.6963+08.3743 &                &                &                &                &                &                & $4.33\pm 0.12$ &                 \\ 
Pristine\,254.0070+12.7611 & $6.15\pm 0.07$ & $6.61\pm 0.16$ &                & $7.48\pm 0.12$ & $7.63\pm 0.03$ &                & $6.24\pm 0.17$ & $3.49\pm 0.14$  \\ 
Pristine\,254.1519+12.6741 & $6.00\pm 0.13$ & $6.23\pm 0.20$ &                & $7.24\pm 0.08$ &                &                & $6.09\pm 0.19$ & $3.14\pm 0.20$  \\ 
Pristine\,254.4842+15.4573 & $6.26\pm 0.26$ & $6.47\pm 0.20$ &                & $7.54\pm 0.05$ & $7.81\pm 0.48$ &                & $6.34\pm 0.04$ & $3.30\pm 0.13$  \\
Pristine\,254.5606+15.4784 & $5.83\pm 0.13$ & $5.91\pm 0.20$ &                & $7.10\pm 0.12$ &                &                & $5.69\pm 0.20$ & $2.68\pm 0.18$  \\
Pristine\,255.0531+10.7488 & $6.20\pm 0.00$ & $6.52\pm 0.20$ &                & $7.76\pm 0.32$ &                &                &                & $3.17\pm 0.21$  \\ 
\hline
\end{tabular}
\end{table*}

\begin{table*}
\caption{Stellar individual abundances from Ti to Fe as: 
$A\left({\rm X}\right)=\log_{10}\left({N\left({\rm X}\right)\over N\left({\rm H}\right)}\right)+12$.}
\renewcommand{\tabcolsep}{3pt}
\tabskip=0pt
\footnotesize
\label{tlab_abun2}
\begin{tabular}{lccccccc}\hline
Star & \ion{Ti}{i} & \ion{Ti}{ii} & \ion{V}{i} & \ion{Cr}{i} & \ion{Mn}{i} & \ion{Fe}{i} & \ion{Fe}{ii} \\
\hline 
Pristine\,180.6670+13.3324 & $4.98\pm 0.24$ & $4.78\pm 0.17$ &                & $5.31\pm 0.04$ &                & $6.92\pm 0.17$ & $6.90\pm 0.24$ \\ 
Pristine\,180.8913+11.3199 & $4.82\pm 0.18$ & $4.61\pm 0.32$ &                & $5.32\pm 0.21$ & $4.38\pm 0.30$ & $7.01\pm 0.18$ & $7.02\pm 0.21$ \\ 
Pristine\,185.0799+14.6464 & $4.26\pm 0.40$ & $3.73\pm 0.09$ &                & $4.43\pm 0.00$ & $3.77\pm 0.30$ & $6.04\pm 0.31$ & $6.04\pm 0.40$ \\ 
Pristine\,194.8481+11.5875 &                &                &                &                & $3.61\pm 0.30$ & $4.88\pm 0.31$ & $4.92\pm 0.30$ \\ 
Pristine\,196.6013+15.6768 & $5.01\pm 0.23$ & $4.82\pm 0.25$ & $4.15\pm 0.07$ & $5.47\pm 0.08$ & $5.07\pm 0.30$ & $7.20\pm 0.22$ & $7.18\pm 0.19$ \\ 
Pristine\,196.9043+06.8973 & $4.89\pm 0.28$ & $5.02\pm 0.04$ &                & $5.51\pm 0.11$ & $5.36\pm 0.30$ & $7.45\pm 0.16$ & $7.45\pm 0.17$ \\ 
Pristine\,205.4970+15.0564 & $3.72\pm 0.30$ & $3.41\pm 0.19$ & $3.60\pm 0.00$ &                & $3.75\pm 0.30$ & $5.93\pm 0.42$ & $5.88\pm 0.39$  \\
Pristine\,210.6952+12.8768 & $4.60\pm 0.19$ & $4.76\pm 0.01$ & $3.52\pm 0.16$ & $5.66\pm 0.00$ & $4.92\pm 0.30$ & $7.10\pm 0.23$ &                \\ 
Pristine\,220.6472+15.7418 & $5.05\pm 0.30$ & $5.02\pm 0.17$ & $4.22\pm 0.30$ & $5.43\pm 0.13$ & $4.59\pm 0.30$ & $7.15\pm 0.24$ & $7.08\pm 0.24$ \\ 
Pristine\,224.8486+07.0259 & $5.05\pm 0.14$ & $5.10\pm 0.17$ & $4.07\pm 0.08$ & $5.64\pm 0.20$ & $5.25\pm 0.04$ & $7.38\pm 0.14$ & $7.37\pm 0.14$  \\
Pristine\,230.4662+06.5251 & $4.66\pm 0.23$ & $4.45\pm 0.30$ & $3.43\pm 0.00$ & $4.96\pm 0.03$ &                & $6.74\pm 0.20$ & $6.73\pm 0.22$  \\
Pristine\,231.0319+06.4867 & $5.05\pm 0.12$ & $5.08\pm 0.28$ & $4.01\pm 0.17$ & $5.84\pm 0.14$ & $5.42\pm 0.06$ & $7.55\pm 0.16$ & $7.54\pm 0.23$  \\ 
Pristine\,231.2818+06.4018 & $4.78\pm 0.17$ & $4.85\pm 0.17$ & $3.69\pm 0.12$ & $5.63\pm 0.18$ & $5.15\pm 0.12$ & $7.38\pm 0.12$ & $7.38\pm 0.20$  \\
Pristine\,233.5730+06.4702 &                &                &                &                &                & $5.24\pm 0.24$ & $5.44\pm 0.30$  \\
Pristine\,237.0863+10.5790 & $4.93\pm 0.22$ & $4.69\pm 0.07$ & $4.07\pm 0.16$ & $5.73\pm 0.16$ & $5.25\pm 0.30$ & $7.43\pm 0.17$ & $7.43\pm 0.18$ \\ 
Pristine\,245.8356+13.8777 & $3.49\pm 0.30$ & $3.42\pm 0.20$ &                &                &                & $5.40\pm 0.33$ & $5.33\pm 0.06$ \\ 
Pristine\,249.2044+10.5327 & $3.35\pm 0.13$ & $3.12\pm 0.23$ &                & $3.54\pm 0.00$ & $2.68\pm 0.30$ & $5.66\pm 0.20$ & $5.66\pm 0.31$  \\
Pristine\,250.6963+08.3743 &                &                &                &                &                & $5.40\pm 0.22$ & $5.71\pm 0.00$  \\ 
Pristine\,254.0070+12.7611 & $5.00\pm 0.13$ & $4.93\pm 0.23$ & $4.18\pm 0.30$ & $5.56\pm 0.34$ & $5.37\pm 0.28$ & $7.35\pm 0.18$ & $7.34\pm 0.14$  \\ 
Pristine\,254.1519+12.6741 & $4.83\pm 0.08$ & $4.91\pm 0.09$ & $3.90\pm 0.03$ & $5.52\pm 0.15$ & $5.00\pm 0.14$ & $7.31\pm 0.16$ & $7.31\pm 0.05$  \\ 
Pristine\,254.4842+15.4573 & $4.89\pm 0.15$ & $4.99\pm 0.14$ & $3.85\pm 0.03$ & $5.68\pm 0.11$ & $5.34\pm 0.03$ & $7.44\pm 0.14$ & $7.44\pm 0.13$  \\
Pristine\,254.5606+15.4784 & $4.34\pm 0.26$ & $4.56\pm 0.15$ & $3.35\pm 0.21$ & $4.70\pm 0.30$ & $4.65\pm 0.33$ & $6.85\pm 0.23$ & $6.86\pm 0.30$  \\
Pristine\,255.0531+10.7488 & $4.66\pm 0.31$ & $4.90\pm 0.12$ & $3.77\pm 0.14$ & $5.66\pm 0.31$ &                & $7.35\pm 0.21$ &                 \\ 
\hline
\end{tabular}
\end{table*}

\begin{table*}
\caption{Stellar individual abundances from Co to Ba as: 
$A\left({\rm X}\right)=\log_{10}\left({N\left({\rm X}\right)\over N\left({\rm H}\right)}\right)+12$.}
\renewcommand{\tabcolsep}{3pt}
\tabskip=0pt
\footnotesize
\label{tlab_abun3}
\begin{tabular}{lcccccc}\hline
Star & \ion{Co}{i} & \ion{Ni}{i} & \ion{Cu}{i} & \ion{Zn}{i} & \ion{Y}{ii} & \ion{Ba}{ii} \\
\hline 
Pristine\,180.6670+13.3324 &                & $5.89\pm 0.19$ & $4.02\pm 0.30$ & $3.85\pm 0.30$ &                &  $2.03\pm 0.30$ \\ 
Pristine\,180.8913+11.3199 &                & $5.91\pm 0.18$ &                & $3.79\pm 0.30$ & $1.53\pm 0.30$ &  $2.36\pm 0.30$ \\ 
Pristine\,185.0799+14.6464 &                &                &                &                &                &  $0.48\pm 0.30$ \\ 
Pristine\,194.8481+11.5875 &                &                &                &                &                &  $0.36\pm 0.69$ \\ 
Pristine\,196.6013+15.6768 &                & $6.07\pm 0.20$ & $3.91\pm 0.17$ &                & $1.90\pm 0.30$ &  $2.02\pm 0.30$ \\ 
Pristine\,196.9043+06.8973 & $5.38\pm 0.20$ & $6.40\pm 0.16$ &                &                & $1.88\pm 0.27$ &                \\ 
Pristine\,205.4970+15.0564 &                &                &                & $2.91\pm 0.30$ &                & $-0.19\pm 0.30$ \\
Pristine\,210.6952+12.8768 &                & $5.92\pm 0.20$ &                &                &                &  $1.00\pm 0.30$ \\ 
Pristine\,220.6472+15.7418 &                & $5.91\pm 0.33$ &                &                &                &                \\ 
Pristine\,224.8486+07.0259 & $5.08\pm 0.04$ & $6.23\pm 0.11$ & $4.51\pm 0.30$ & $4.12\pm 0.30$ & $1.79\pm 0.30$ &  $1.69\pm 0.30$ \\
Pristine\,230.4662+06.5251 &                & $5.65\pm 0.33$ &                & $4.05\pm 0.30$ & $1.48\pm 0.30$ &                 \\
Pristine\,231.0319+06.4867 & $5.09\pm 0.07$ & $6.46\pm 0.26$ &                & $4.79\pm 0.30$ & $2.14\pm 0.30$ &  $2.86\pm 0.30$ \\ 
Pristine\,231.2818+06.4018 & $4.79\pm 0.20$ & $6.28\pm 0.14$ & $4.10\pm 0.30$ &                & $2.03\pm 0.30$ &  $2.03\pm 0.30$ \\
Pristine\,233.5730+06.4702 &                &                &                &                &                & $-0.47\pm 0.30$ \\
Pristine\,237.0863+10.5790 &                & $6.21\pm 0.18$ & $4.24\pm 0.30$ &                & $1.73\pm 0.30$ &  $2.41\pm 0.30$ \\ 
Pristine\,245.8356+13.8777 &                &                &                &                &                &                \\ 
Pristine\,249.2044+10.5327 &                &                &                & $2.88\pm 0.30$ &                &  $0.17\pm 0.03$ \\
Pristine\,250.6963+08.3743 &                &                &                &                &                & $-0.26\pm 0.30$ \\ 
Pristine\,254.0070+12.7611 &                & $6.26\pm 0.18$ &                &                & $1.95\pm 0.06$ &  $1.76\pm 0.30$ \\ 
Pristine\,254.1519+12.6741 &                & $6.23\pm 0.33$ & $3.95\pm 0.01$ & $3.89\pm 0.30$ & $1.14\pm 0.45$ &  $2.04\pm 0.30$ \\ 
Pristine\,254.4842+15.4573 & $4.86\pm 0.08$ & $6.35\pm 0.26$ & $4.43\pm 0.30$ & $4.69\pm 0.30$ & $2.00\pm 0.06$ &  $1.97\pm 0.30$ \\
Pristine\,254.5606+15.4784 & $4.61\pm 0.05$ & $5.48\pm 0.33$ &                &                & $1.47\pm 0.30$ &  $1.10\pm 0.34$ \\
Pristine\,255.0531+10.7488 & $5.02\pm 0.20$ & $6.06\pm 0.05$ & $4.01\pm 0.30$ &                &                &                 \\ 
\hline
\end{tabular}
\end{table*}

\section{Discussion}

In the course of the spectroscopic follow-up of the Pristine survey, that currently
covers 1000 deg$^2$ \citep{pristine1}, the analysis of this sample of bright stars has 
revealed that most of the stars are not as metal-poor as expected and this is a direct consequence of the inadequacy 
of the SDSS photometry for these bright sources.
The Pristine selection criteria has been updated since these stars were observed
and now the success rate is much higher (see Youakim et al. 2017 submitted).
Recently, the Pristine collaboration performed a detailed analysis of 
the selection criteria used for choosing spectroscopic follow-up targets 
from Pristine and SDSS photometry (Youakim et al. 2017, submitted). 
Applying these criteria to the sample here analysed, eliminates half of the stars,
most of which are metal-rich. 
The stars selected by the new calibration are labelled (Y) in Table 2.
All but one of the remaining stars are predicted by Pristine + SDSS photometry 
to have $\mathrm[Fe/H] \leq -2.0$, but our current spectroscopic analysis confirms 
only 5 of these stars to actually have $\mathrm[Fe/H] \leq -2.0$. 
This leaves 5/10 (50 \%) stars in the current sample as contaminants — a significantly 
higher contaminant fraction than the 8\% reported in Youakim et al. (2017, submitted). 
We can identify two of these as cool stars (${\rm T}_{\rm eff} < 4700$\,K) 
that lie on the edge of the SDSS $g-i$ colour space for which Pristine can reliably 
derive photometric metallicities, but nonetheless the contamination rate is still higher than expected. 
We note that the current sample is small, and that more follow-up spectroscopy of bright stars 
is needed to investigate this further. Should this persist as a larger sample becomes available, 
this may suggest that at bright magnitudes the SDSS photometry still reaches saturation in the $g-$, $r-$, 
or $i-$ bands despite not having been flagged as such. In either case, this reaffirms our findings 
that APASS is more appropriate than SDSS to use with Pristine for brighter targets.
Figures \ref{plotgi} and \ref{plotgr} show that the measured metallicities are
consistent with the colours obtained by combining Pristine photometry with APASS $gri$.
Given the narrow band of the CaHK filter, its bright end ($CaHK$=9) is considerably brighter than the saturation limit of SDSS.
At these magnitudes APASS instead provides accurate photometry that is the ideal complement to the Pristine photometry.  

Also for stars that are bright enough for $gri$ magnitudes to be saturated 
the $u$ magnitudes are often not saturated. This is due to a combination
of the low efficiency of the SDSS $u$ filter, atmospheric extinction
and decreasing UV flux for F, G, K stars that are our scientific target. 
In Fig.\,\ref{ug_gr} we show our stars in the $(u-g)_0$ vs.
$(g-r)_0$ diagram in which all the bands are APASS except for $u$,
that is taken from SDSS. This diagram is a powerful diagnostic
of atmospheric parameters, as shown by the overlayed synthetic colours diagnostic.

In principle the leverage on metallicity that is afforded by
the CaHK filter  should also allow to use the $u-g$ colour
as a luminosity indicator, at least to the point of discriminating dwarf stars from giants.

However, inspection of Figures  \ref{ug_cak_m00} and \ref{ug_cak_m05} shows that the diagnostic does not work
as well as expected. On the one hand we can question the accuracy of the $(u-g)_0$ colour 
obtained by combining SDSS $u$ and APASS $g$, some calibration
might be necessary; on the other hand it is useful to recall that
our modelling of the stellar atmospheres may be unable to simultaneously model
the iron ionisation equilibrium and the Balmer jump ($u-g$ colour).  
Further investigation of these issues is needed to properly asses
to which level Pristine photometry, coupled to the $u-g$ colour
may provide a reliable dwarf/giant discrimination.

\section{Conclusions and future perspectives}

The most important conclusion of this investigation is that the
bright end of the Pristine survey can be exploited, if coupled to suitable
broad band photometry. The APASS photometry is providing an ideal 
complement to Pristine, for $g\le 15$.

For this magnitude range, 2\,m class telescopes with efficient spectrographs, 
such as FEROS at the MPG/ESO 2.2\,m telescope, are suitable for high resolution follow-up. 

We shall cross-identify the Pristine survey with APASS and use the
two together in order to identify metal-poor and extremely metal-poor stars.
These bright targets will be the object of spectroscopic
follow-up using FEROS and other spectrographs on 2\,m class telescopes.
Instruments like FEROS have, typically, low efficiency in the UV and in the
IR, so that the strongest metallic lines are not available. 
Ruling out the \ion{Ca}{ii} H\&K and IR triplet, the most readily available
metallic features, for EMP stars, are the \ion{Mg}{i} b triplet and the 
G-band. These features are much stronger in K giants than in F dwarfs.
For this reason we shall give special attention to K giants, 
leaving F dwarfs to larger telescopes in the follow-up with 2\,m class telescopes.

The plan is to make available to the community all
the reduced spectra, as well as spectroscopic and photometric catalogues, as soon
as human resources for this effort will become available.

\acknowledgements
We are grateful to the referee G. Bono for helping us to improve the paper.
This paper makes use of data from the AAVSO Photometric All Sky Survey, 
whose funding has been provided by the Robert Martin Ayers Sciences Fund.
We acknowledge support from CNRS/INSU through PICS grant.
ES and KY gratefully acknowledge funding by the Emmy Noether program from the Deutsche Forschungsgemeinschaft (DFG).
APASS is funded through NSF grant AST-1412587.
DA acknowledges the Spanish Ministry of Economy and Competitiveness (MINECO) 
for the financial support received in the form of a Severo-Ochoa PhD fellowship, 
within the Severo-Ochoa International Ph.D. Pro- gram. 
JIGH, DA, and CAP also 
acknowledge the Spanish ministry project MINECO AYA2014-56359-P. 
JIGH acknowledges financial support from the Spanish Ministry of Economy and 
Competitiveness (MINECO) under the 2013 Ramo ́n y Ca jal program MINECO RYC-2013-14875.

\end{document}